\documentclass{PoS}
\usepackage{amssymb,amsmath,bm}
\DeclareMathOperator{\tr}{Tr}
\renewcommand{\d}{\mathrm{d}}
\newcommand{\qT}{\bm{q}_\sT}
\newcommand{\sT}{{\scriptscriptstyle T}}
\newcommand{\xB}{x_{\scriptscriptstyle B}}
\newcommand{\mc}[1]{\mathcal{#1}}

\title{Momentum Imbalance Observables as a Probe of Gluon TMDs}

\ShortTitle{Momentum Imbalance Observables as a Probe of Gluon TMDs}

\author{\speaker{Cristian Pisano} \\
        Department of Physics, University of Antwerp, Groenenborgerlaan 171, 2020 Antwerp, Belgium\\
        E-mail: \email{cristian.pisano@uantwerp.be}}

\abstract{The unpolarized and linearly polarized gluon TMDs can be directly probed in heavy quark and jet pair production in unpolarized electron-proton collisions by looking at observables, like transverse momentum distributions and azimuthal asymmetries, depending on the momentum imbalance of the pair. 
Analytical expressions are presented for these observables and for analogous ones in Higgs plus jet and quarkonium plus photon production in unpolarized proton-proton scattering experiments. It is shown how the proposed measurements, to be performed at a future EIC and at the LHC, could provide important information  on the size and shape of gluon TMDs, as well as on other fundamental properties such as their process and energy scale dependences.}

\FullConference{QCD Evolution 2015 -QCDEV2015-\\
		26-30 May 2015\\
		Jefferson Lab (JLAB), Newport News Virginia, USA}

\begin{document}

\section{Introduction}

Transverse momentum dependent parton distribution functions (TMDs) give valuable insight into the intrinsic motion of partons and the correlation between their spins and momenta, providing a full three-dimensional picture of hadrons in momentum space. Their knowledge is necessary to understand the observed polarization effects in high-energy scattering processes, like transverse single spin asymmetries~\cite{asym1,asym2,asym3}, which have gathered considerable attention from both experimental and theoretical communities in recent years. From the theory point of view, proofs of TMD factorization exist for a few processes in which two energy scales are present: a hard one, $Q$,  and a transverse momentum, $q_\sT$, such that $q_\sT\ll Q$. Typical examples are semi-inclusive deeply inelastic scattering (SIDIS), Drell-Yan (DY) and $e^+e^-$ annihilation processes~\cite{Collins:1985ue,Ji:2004xq,Collins:2011zzd,GarciaEchevarria:2011rb}. On the experimental side, much effort has been made to extract the \emph{quark} TMDs from low energy data, mainly from HERMES, COMPASS and JLab measurements~\cite{asym1,asym2}. On the contrary, almost nothing is known experimentally about \emph{gluon} TMDs~\cite{Boer:2011fh,pages:2015ika}.

If the proton is unpolarized, in addition to the unpolarized gluon TMD denoted by $f_1^g$, a distribution of linearly polarized gluons, $h_1^{\perp \,g}$, corresponding to an interference between $+1$ and $-1$ gluon helicity states, can in principle be nonzero~\cite{Mulders:2000sh}. In Ref.~\cite{Boer:2009nc} it was found that $h_1^{\perp \,g}$ can contribute to the dijet imbalance in hadronic collisions, commonly used to determine the average transverse momentum squared of partons inside protons. Depending on its size and on whether its contribution can be calculated and taken into account, $h_1^{\perp \,g}$ can complicate or even hamper this extraction. Hence it is important to determine its size separately,  using other observables. Although in Ref.~\cite{Boer:2009nc}  a method has been discussed to isolate  $h_1^{\perp \,g}$ by means of an azimuthal angular dependent weighting of the cross section,  such a process is expected to suffer from contributions that break factorization, through initial (ISI) and final state interactions (FSI)~\cite{Rogers:2010dm}. 

In the present contribution, after providing a formal definition of gluon TMDs in QCD, heavy quark pair and dijet production in electron-proton collisions is discussed. Such processes, that could be studied at a future Electron Ion Collider (EIC), offer a theoretically clean and safe way to probe both $f_1^g$ and $h_1^{\perp\,g}$ because TMD factorization is expected to hold~\cite{Boer:2010zf,Pisano:2013cya} . Furthermore, it is shown how complementary information could come, in proton-proton collisions, from the analysis of Higgs boson production in association with a jet~\cite{Boer:2014lka} and from quarkonium production in association with a photon~\cite{Dunnen:2014eta}. For all these process, estimates are provided for observables sensitive to gluon TMDs, which depend on the transverse momentum imbalance of two particles or jets in the final state.

\section{Operator definition of gluon TMDs}
The information on the gluon TMDs of an unpolarized proton carrying four-momentum $P$ is encoded in the transverse momentum dependent correlator, defined as follows~\cite{Mulders:2000sh}
\begin{eqnarray}
\label{GluonCorr}
\Gamma_g^{[U]\,\mu\nu}(x,\bm p_\sT )
& = &  \frac{n_\rho\,n_\sigma}{(p{\cdot}n)^2}
{\int}\frac{\d(\xi{\cdot}P)\,\d^2\xi_\sT}{(2\pi)^3}\
e^{ip\cdot\xi}\,
\langle P|\,\tr_c\big[\,F^{\mu\rho}(0)\,U_{[0,\xi]}\,
F^{\nu\sigma}(\xi)\,U^{\prime}_{[\xi,0]}\,\big]
\,|P \rangle\,\big\rfloor_{\text{LF}} \nonumber \\
& =& 
-\frac{1}{2x}\,\bigg \{g_\sT^{\mu\nu}\,f_1^{g\, [U]} (x,\bm p_\sT^2)
-\bigg(\frac{p_\sT^\mu p_\sT^\nu}{M_p^2}\,
{+}\,g_\sT^{\mu\nu}\frac{\bm p_\sT^2}{2M_p^2}\bigg)
\;h_1^{\perp\,g \,[U]} (x,\bm p_\sT^2) \bigg \} , \label{Phig}
\label{eq:corrg}
\end{eqnarray}
where $n$ is a lightlike vector conjugate to $P$, $p$ is the gluon momentum decomposed as $p = x P +p_\sT + p^-n $ (with  $p_\sT^2 = - \bm p_\sT^2$) and $F^{\mu\nu}$ denotes the gluon field strength. Moreover, a trace over color is taken in the integrand, where non-locality is limited to the lightfront (LF), $\xi \cdot n  = 0$. The process dependent gauge links $U_{[0,\xi]}$ and $U^{\prime}_{[\xi,0]}$ are path ordered exponentials in the triplet representation and are needed to ensure the gauge invariance of the correlator.

Since the above correlator cannot be calculated from first principles, its expansion in terms of TMDs is commonly
 used for phenomenological applications. This expansion is given at the leading twist level, with the naming convention of Ref.~\cite{Meissner:2007rx}, in the second line of Eq.~(\ref{eq:corrg}), where we have introduced the transverse projector $g^{\mu\nu}_{\sT} = g^{\mu\nu} - P^{\mu}n^{\nu}/P\cdot n -n^{\mu}P^{\nu}/P\cdot n$.  The function $f_1^{g\,[U]}$ is the distribution of unpolarized gluons inside an unpolarized proton, while  $h_1^{\perp\,g\,[U]}$  represents the $T$-even distribution of linearly polarized gluons inside an unpolarized proton. The function
 $h_1^{\perp\,g\,[U]}$ is also even in $p_\sT$ because it enters in the correlator as a
rank 2 tensor, describing a $\Delta L = 2$ helicity-flip distribution. Furthermore, it satisfies the following, model-independent, positivity bound,
 \begin{equation}
 \frac{\vert \bm p_\sT^2 \vert}{2 M_p^2}\, \vert h_{1\sT}^{\perp \,g\,[U]}(x,\bm p_\sT^2) \vert  \le   
 f_1^{g\,[U]} (x,\bm p_\sT^2)~.
 \label{eq:bound}
 \end{equation}

Like all other TMDs, $f_1^{g\, [U]}$ and $h_1^{\perp\,g\,[U]}$ receive contributions from ISI or FSI,  summed up into the gauge links, that lead to their process dependence or even to violations of QCD factorization, as already mentioned above. It is therefore important to consider extractions from several different processes, like the ones proposed in the following. In this way,  the nonuniversality and factorization breaking issues can be properly studied and quantified.

\section{Heavy quark pair and dijet production in DIS}
Assuming TMD factorization, the cross section for the process $e (\ell)+p(P)\to e(\ell')+ Q(K_1) + \bar{Q}(K_2)+X$,
with the four-momenta of the particles given inside brackets, can be written as 
\begin{align}
\d\sigma
& = \frac{1}{2 s}\,\frac{\d^3 \ell'}{(2\pi)^3\,2 E_e^{\prime}} \frac{\d^3 K_{Q}}{(2\pi)^3\,2 E_{Q}}
\frac{\d^3K_{\bar Q}}{(2\pi)^3\,2E_{\bar Q}}
{\int}\d x\, \d^2\bm p_{\sT}\,(2\pi)^4
\delta^4(q {+} p {-} K_{Q} {-} K_{\bar Q})
 \nonumber \\
&\qquad \qquad\qquad \times \sum_{a, b, c}\ \frac{1}{Q^4}\, 
\tr \bigg\{L(\ell,q) \otimes\Gamma_a(x{,}\bm p_{\sT})
\otimes |H_{\gamma^*\, a \rightarrow b\, c }(q, p, K_{Q}, K_{\bar Q})|^2 \bigg \}\, , 
\label{eq:CrossSec}
\end{align}
where $s= (\ell+P)^2$, $q= \ell-\ell^\prime$ is the four-momentum of the exchanged virtual photon with $q^2=-Q^2$, and the symbol $\otimes$  denotes appropriate traces over the Lorentz and Dirac indices. The sum runs over all the partons in the initial and final states, while the leptonic tensor $L(\ell, q) $ is given by
\begin{equation}
L^{\mu\nu}(\ell, q) = -g^{\mu\nu}\,Q^2 + 2\,( 
 \ell^{\mu}\ell'^\nu + \ell^{\nu}\ell'^\mu)\,,
\end{equation}
and $H_{\gamma^*a\to bc}$ is the amplitude for the partonic hard subprocess $\gamma^*a \to  bc$. At leading order (LO) in perturbative QCD, only the channel $\gamma^*g\to Q \bar Q$ contributes. By substituting the parameterization of the gluon correlator, Eq.~(\ref{eq:corrg}), into Eq.~(\ref{eq:CrossSec}), one obtains, in the 
$\gamma^*$-$P$ center-of-mass frame and in the specific kinematic configuration in which the quark-antiquark pair is almost back-to-back in the 
plane orthogonal to the direction of $P$ and $\gamma^*$,
\begin{align}
\!\!\frac{\d\sigma}{\d y_Q\,\d y_{\bar Q}\,\d y\,\d x_{\xB}\,\d^2\bm{q}_{\sT} \d^2\bm{K}_{\perp}} & = 
\frac{\alpha^2\alpha_s}{\pi s M_\perp^2}\, \frac{1}{\xB y^2}\, 
\bigg\{ A_0 + A_{1}\cos \phi_\perp + A_{2} \cos 2 \phi_\perp   \nonumber \\
& \quad +\, \bm q_\sT^2 \, \left [B_0 \cos 2 (\phi_\perp-\phi_\sT)
+\, B_{1} \cos (\phi_\perp-2\phi_\sT) \,+\,   B^{\prime}_{1} \cos (3 \phi_\perp-2\phi_\sT) \right . \nonumber \\
& \quad \quad +\, B_{2} \cos 2 \phi_\sT  + \,\left . B^{\prime}_{2} \cos 2(2\phi_\perp-\phi_\sT) \right ] \bigg\}\,\delta(1-z_1-z_2)~.
\label{eq:cscomplete}
\end{align}
In Eq.~(\ref{eq:cscomplete}),  $y_{Q/\bar Q}$ are the rapidities of the final heavy quark/antiquark with mass $M_Q$,  $z_{1/2} = P~\cdot~K_{Q/\bar Q}/P \cdot~q$, $y= P \cdot q/P \cdot \ell$ and $x_B$ is the Bjorken variable. Moreover, we have introduced the sum and difference of the transverse heavy quark momenta, $K_\perp = (K_{Q\perp} - K_{\bar Q\perp})/2$ and $q_\sT = K_{Q\perp}+ K_{\bar Q\perp}$ with $\vert q_\sT \vert \ll \vert K_\perp \vert$. Therefore, we use the approximate transverse momenta $K_{Q\perp}\approx K_\perp$, 
$K_{\bar Q\perp} \approx -K_\perp$, and $M_{Q\perp}^2 \approx M_{\bar Q\perp}^2 \approx M^2_\perp = M_Q^2 + \bm K^2_\perp$. The azimuthal
angles of $\bm q_\sT$ and $\bm K_\perp$, denoted as $\phi_\sT$ and $\phi_\perp$ respectively, are measured w.r.t.\ the leptonic plane $(\phi_\ell=\phi_{\ell^\prime})$.  

\begin{figure}[t]
\centering
\includegraphics[width=5.6cm]{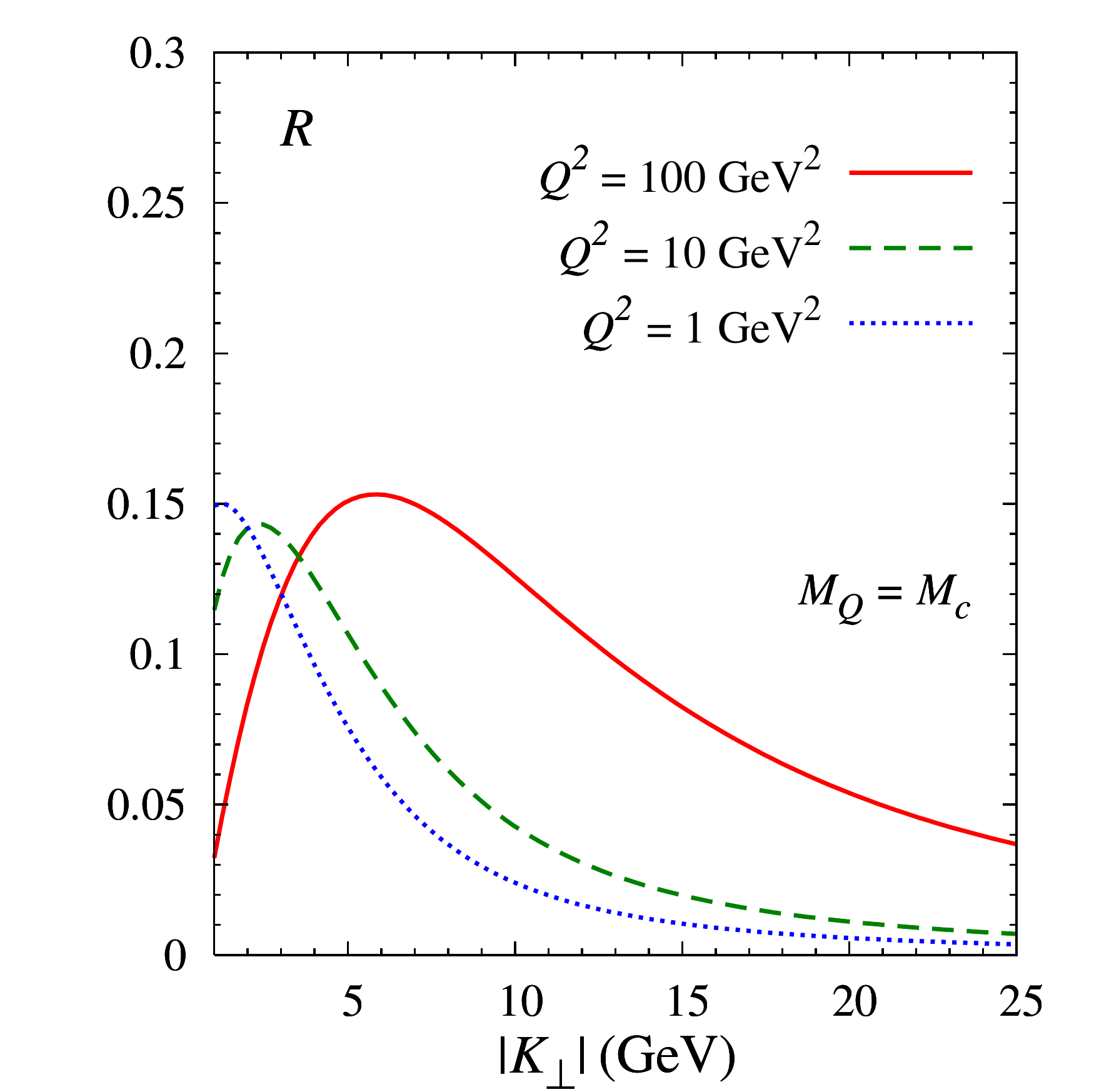}\hspace{0.5cm}
\includegraphics[width=5.6cm]{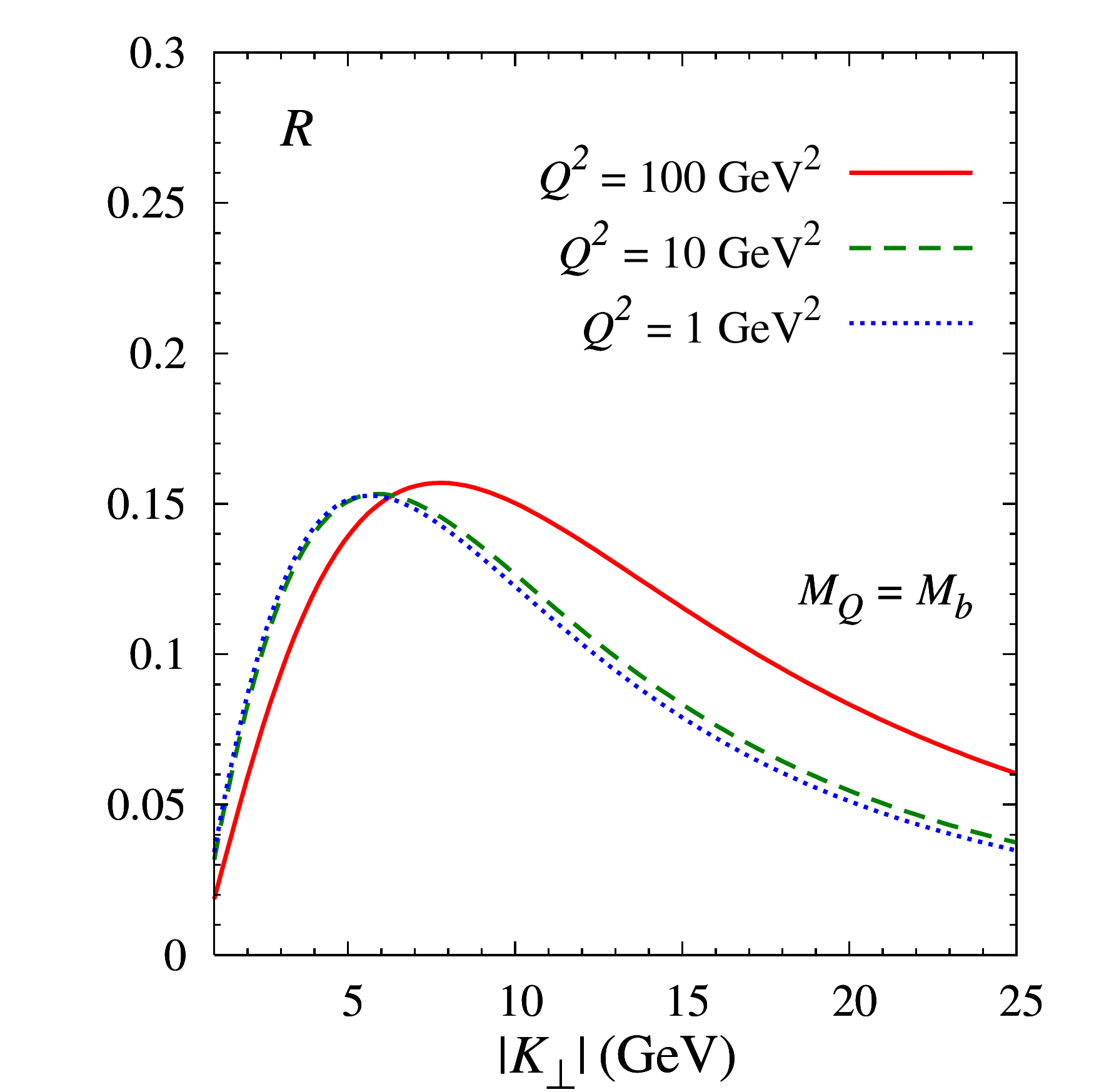}
\caption{Upper bounds $R$ on $\vert \langle \cos 2(\phi_\perp -\phi_\sT )\rangle\vert$ for the process 
$ep \to  e^\prime Q \bar Q X$, as a function of 
$\vert \bm K_\perp \vert $ at $z_1 = z_2=1/2$, $y = 0.01$ and different values of $Q^2$, for charm (left panel) and bottom (right panel) production.}
\label{fig:asy}
\end{figure}

The terms $A_i$  in Eq.~(\ref{eq:cscomplete}) depend only on  one TMD, $f_1^g$, while $B_i^{(\prime)}$ depend only on $h_1^{\perp\,g}$, and their explicit expressions can be found in Ref.~\cite{Pisano:2013cya}. The different azimuthal modulations in Eq.~(\ref{eq:cscomplete}) can be singled out by defining the following weighted cross sections
\begin{equation}
\langle W(\phi_\perp, \phi_\sT) \rangle \equiv  \frac{\int 
\d \phi_\perp\d \phi_\sT
\, W(\phi_\perp, \phi_\sT) \, \d\sigma}{\int \d \phi_\perp \d \phi_\sT\,
\, \d\sigma}\,,
\end{equation}
with $W(\phi_\perp, \phi_\sT)$ being one of the circular functions of $\phi_\perp$ and $\phi_\sT$ in Eq.~(\ref{eq:cscomplete}).

\begin{figure}[t]
\centering
\includegraphics[width=5.6cm]{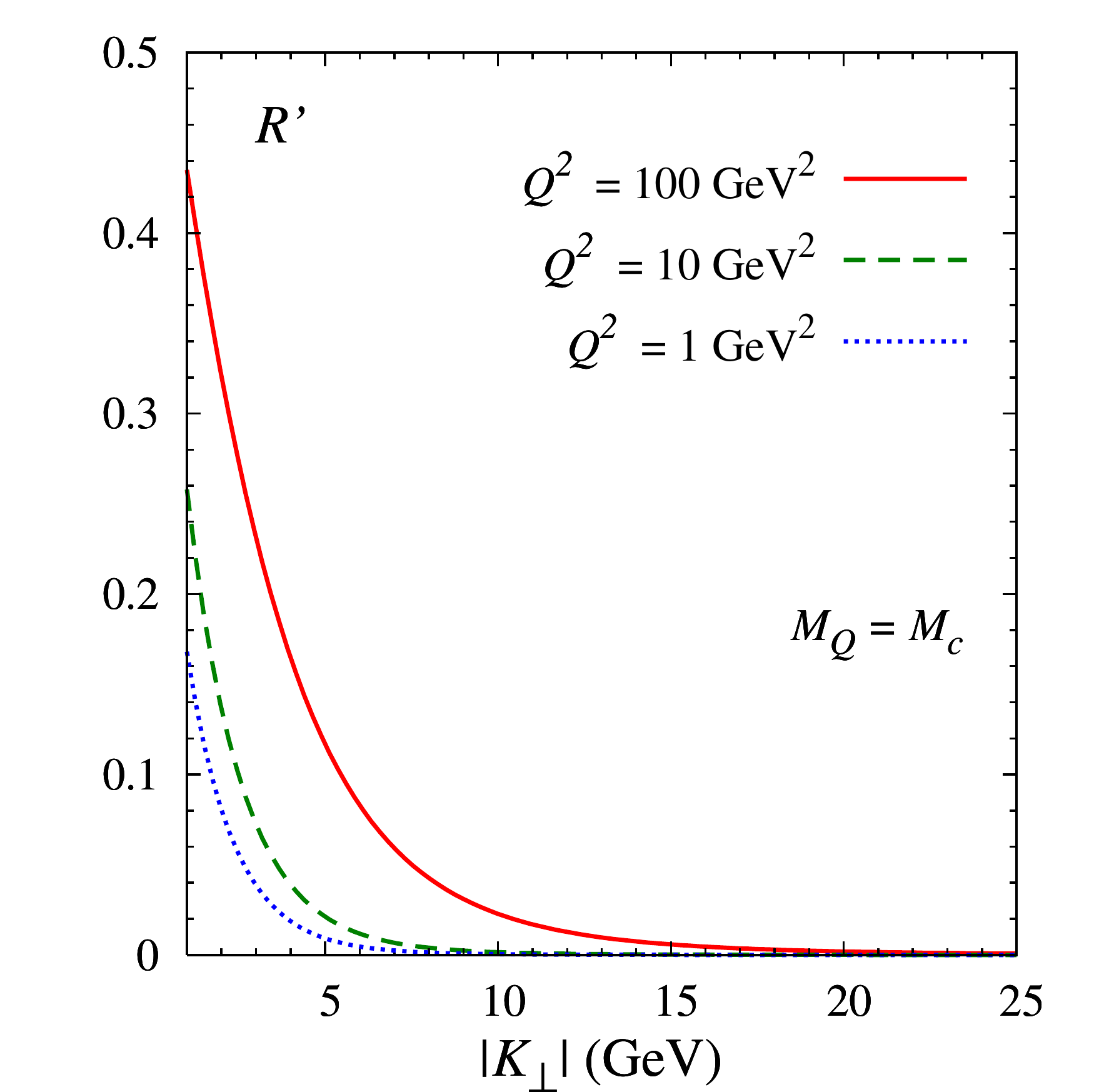}\hspace{0.5cm}
\includegraphics[width=5.6cm]{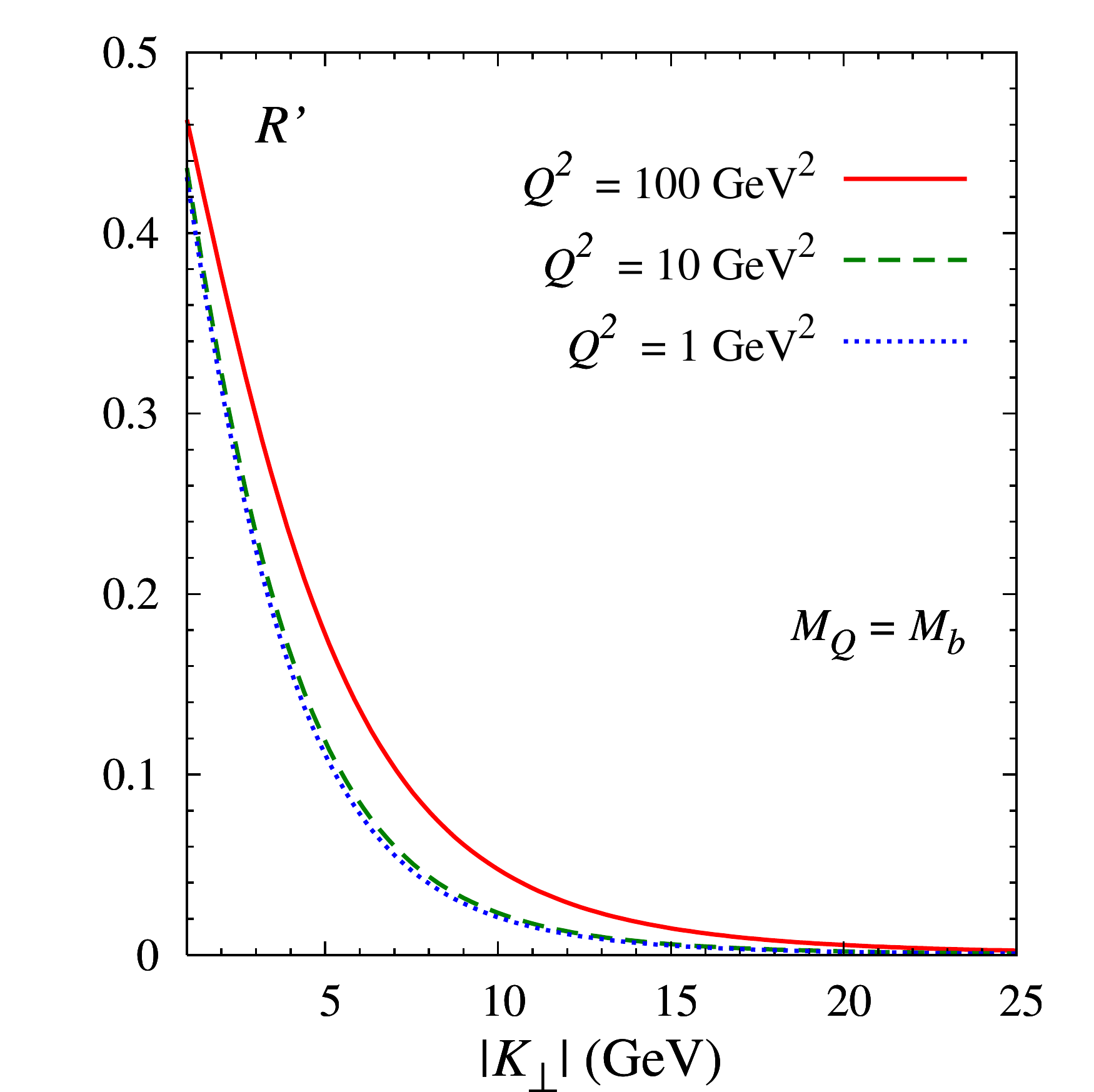}
\caption{Same as in Fig.~1, but for the upper bound $R^\prime$ on $\cos2\phi_\sT$}
\label{fig:asyp}
\end{figure}

By means of the positivity bound in Eq.~(\ref{eq:bound}),  one can determine the maximum values of the asymmetries originating from the linear polarization of gluons. For example,  the upper bound $R$ on $\vert \langle \cos 2(\phi_\perp - \phi_\sT )\rangle \vert $ is defined as follows:
\begin{equation}
\vert \langle \cos 2 (\phi_\perp-\phi_\sT) \rangle \vert = \left| \frac{\int 
\d \phi_\perp\d \phi_\sT
\, \cos 2 (\phi_\perp-\phi_\sT) \, \d\sigma}{\int \d \phi_\perp \d \phi_\sT
\, \d\sigma}\right| = \frac{ \bm{q}_\sT^2\, | B_0|}{
2 \, A_0} =  \frac{\bm q_\sT^2}{2 M^2}\,\frac{|h_1^{\perp\, g}(x,\bm p_\sT^2 )|}{ f_1^{g}(x,\bm p_\sT^2 )} \, R\, \le R \, ,
\label{eq:R}
\end{equation}
and is presented in Fig.~\ref{fig:asy} as a function of $\vert \boldsymbol K_\perp\vert$ ($>$ 1 GeV)  for charm (left panel) and bottom (right panel)
production at different values of $Q^2$. We have selected $y= 0.01$, $z_1=z_2=1/2$ and taken $M_c^2=$ 2 
GeV$^2$, $M_b^2=$ 25 GeV$^2$. Similarly, the upper bound $R^\prime$ on
$|\langle \cos 2 \phi_\sT\rangle |$ is given by 
\begin{equation}
\vert \langle \cos 2 \phi_\sT \rangle \vert = \left| \frac{\int 
\d \phi_\perp\d \phi_\sT
\, \cos 2 \phi_\sT \, \d\sigma}{\int \d \phi_\perp \d \phi_\sT
\, \d\sigma}\right| = \frac{ \bm{q}_\sT^2\, | B_2|}{
2 \, A_0} = \frac{\bm q_\sT^2}{2 M^2}\,\frac{|h_1^{\perp\, g}(x,\bm p_\sT^2 )|}{ f_1^{g}(x,\bm p_\sT^2 )} \,  R^\prime\,\le\, R^\prime \, ,
\end{equation}
and shown in Fig.\ \ref{fig:asyp} in the same kinematic region as Fig.\ \ref{fig:asy}. Note  
that $R^\prime$ becomes larger than $R$ only at small values of $\vert \boldsymbol K_\perp\vert$, but falls off more rapidly than $R$ as $\vert \boldsymbol K_\perp\vert$ increases. The size of these asymmetries, together with the relative simplicity of the proposed measurement that does not require polarized beams,  will likely allow an extraction of $h_1^{\perp\, g}$ at a future EIC. On the other hand, the unpolarized gluon TMD could be easily determined by integrating Eq.~(\ref{eq:cscomplete}) over $\phi_\perp$ and $\phi_\sT$.  

The cross section for dijet production has the same structure as the one in Eq.~(\ref{eq:cscomplete}), although in 
this case the $A_i$ terms will depend on the partonic process $\gamma^*q\to q g$ as well. Hence, the upper bounds on the azimuthal asymmetries can be obtained from those for heavy quark pair production in the limit  $M_Q\to 0$.

\section{Higgs plus jet production at the LHC}

In addition to the EIC, gluon TMDs can be studied at the LHC as well. For example, inclusive Higgs production has been analyzed within the TMD factorization approach in Refs.~\cite{Boer:2013fca,Boer:2011kf} and, including the effects of QCD evolution, in Refs.~\cite{Sun:2011iw,Boer:2014tka,Echevarria:2015uaa}.  It turns out that the impact of gluon polarization on the Higgs transverse momentum distribution is of the order of a few percent, at the energy scale of the Higgs mass $M_H$. Furthermore, such effects are largest when the transverse momentum of the Higgs boson is small, namely a few GeV, {\it i.e.}\ in a region where the cross section is difficult to measure. In this section the associated production of a Higgs boson and a jet~\cite{Boer:2014lka} is discussed. This process offers some additional features compared to inclusive Higgs production, when it comes to probe gluon TMDs. First of all, one can study their scale evolution by tuning the hard scale, identified for example with the invariant mass of
 the Higgs-jet pair. This is not possible in Higgs production, for which the hard scale is fixed to be $M_H$. Moreover, one can define angular modulations that allow to select specific convolutions of TMDs. Finally,  $h_1^{\perp\, g}$ affects the spectrum of the Higgs-jet pair mostly at values of the transverse momentum of the pair as small as a few GeV, but the single transverse momenta of the Higgs boson and the jet can be much larger.

 To LO in perturbative QCD,  the reaction $p(P_A)\,{+}\,p(P_B)\,\to\, H (K_H) \,{+}\, {\rm jet} (K_{\rm j})\, {+}  \,X$, with the Higgs boson and the jet almost back to back in the plane orthogonal to the direction of the initial protons,  proceeds via the following partonic subprocesses: $gg\to Hg$, $gq\to Hq$ and $q \bar{q} \to Hg$. Consider only the channel $gg\to Hg$, which is the dominant one at the LHC energies, and assuming TMD factorization, the cross section is given by, , similarly to  Eq.~(\ref{eq:CrossSec}), 
\begin{eqnarray}
\d\sigma
& = &\frac{1}{2 s}\,\frac{d^3 \bm K_H}{(2\pi)^3\,2 E_{H}}\, \frac{d^3 \bm K_{\rm j}}{(2\pi)^3\,2 E_{\rm j}}\,
{\int} \d x_a \,\d x_b \,\d^2\bm p_{a\sT} \,\d^2\bm p_{b\sT}\,(2\pi)^4
\delta^4(p_a{+} p_b {-}K_H- K_{\rm j})
 \nonumber \\
&&\qquad \qquad \times
{\rm Tr}\, \left \{  \Gamma_g(x_a {,}\bm p_{a \sT}) \otimes \Gamma_g(x_b {,}\bm p_{b \sT}) \otimes 
 \left|{H}_{g g   \to H g} (p_a, p_b, K_H, K_{\rm j})\right|^2\right \}\,,
\label{eq:CrossSec_Hj}
\end{eqnarray}
where $s=(P_A+P_B)^2$. By neglecting  terms suppressed by powers of $\vert \bm q_\sT \vert /  M_\perp $, with $\bm q_\sT = 
\bm K_{H\perp} +\bm K_{{\rm j} \perp}$  and  
$M_\perp = \sqrt{M_H^2 + \bm K_{H \perp}^2}$, the final result in the laboratory frame reads
\begin{eqnarray}
\frac{\d \sigma}{\sigma} & = & \frac{1}{2\pi}\,\sigma_0(\bm q_\sT^2) \,\left [ 1 + R_0(\bm q_\sT^2) + R_2(\bm q_\sT^2) \cos2\phi +  R_4(\bm q_\sT^2) \cos4\phi\right ]\,,
\label{eq:csgg-2}
\end{eqnarray}
where the normalized cross section is given by
\begin{equation}
\frac{\d\sigma}{ \sigma} \equiv
\frac{ \d\sigma}
{\int_0^{q_{\sT \rm max }^2} \d \bm q_\sT^2\int_0^{2 \pi} \d\phi\, \d\sigma}\,, \qquad \text{with}\qquad
\d\sigma \equiv  \frac{\d\sigma}{\d y_H\, \d y_{\rm j} \,\d^2 \bm{K}_{\perp} \,\d^2 \bm q_\sT}~.
\label{eq:qTdist}
\end{equation}
In the definition above, $y_H$ and $y_j$ are, respectively, the rapidities of the produced Higgs boson and jet along the direction of the incoming protons, $\bm K_\perp =  (\bm K_H-\bm K_{\rm j})/2 \approx  \bm K_{H\perp}\approx  -\bm K_{{\rm j}\perp}$, and $\phi$ is the azimuthal angle between $\bm K_\perp$ and $\bm q_\sT$.
Moreover, we have introduced 
\begin{equation}
\sigma_0(\bm q_\sT^2) = \frac{ {\cal C}[f_1^g \, f_1^g ]}{\int_0^{{q^2_{\sT \rm max }}} \d \bm q^2_\sT \, {\cal C}[f_1^g \, f_1^g ] }\,,
\label{eq:sigma0}
\end{equation}
and the convolution of TMDs
\begin{align}
{\cal{C}}[w\, f\, g]  \equiv  \int d^{2}\bm p_{a\sT}\int d^{2}\bm p_{b\sT}\,
\delta^{2}(\bm p_{a\sT}+\bm p_{b\sT}-\bm q_{\sT}) \, w(\bm p_{a\sT},\bm p_{b\sT})\, f(x_{a},\bm p_{a\sT}^{2})\, g(x_{b},\bm p_{b\sT}^{2})\,,
\label{eq:Conv}
\end{align} 
with $x_{a/b} \, =\, \left (M_{\perp}\,e^{\pm y_H} \, + \, \vert \bm K_{{\rm j} \perp} \vert \,
e^{\pm y_{\rm j}} \right )/{\sqrt s}$, up to corrections of order ${\cal O}(1/s)$.  The terms $R_0$, $R_2$, $R_4$ in Eq.~(\ref{eq:csgg-2}) are functions of $\bm q_\sT^2$ as well as of the Mandelstam variables for the subprocess 
$gg\to Hg$:
\begin{eqnarray}
R_0(\bm q_\sT^2)& =& \frac{M_H^4\,\hat s^2}{M_H^8 + \hat s^4 + \hat t^4 + \hat u^4}\, \frac{{\cal C}[w_0^{hh}\,h_1^{\perp\,g} \, h_1^{\perp\,g} ] } {{\cal C}[f_1^g \, f_1^g ]}\,, \label{eq:R0}\nonumber \\
R_2(\bm q_\sT^2)& = & \frac{ \hat t ^2(\hat t + \hat u)^2 -2 M_H^2 \hat u^2 (\hat t + \hat u) + M_H^4 (\hat t^2 + \hat u^2)  }{M_H^8 + \hat s^4 + \hat t^4 + 
\hat u^4 }  \, \frac{{\cal C}[w_2^{fh} \,f_1^g\,h_1^{\perp\,g}]}{{\cal C}[f_1^g \, f_1^g ]}  \,+ \,(x_a \leftrightarrow x_b , \,\hat t \leftrightarrow \hat u )\,,
\label{eq:R2}  \nonumber  \\
R_4(\bm q_\sT^2) & = & \frac{\hat t^2 \hat u^2}{M_H^8+\hat s^4 + \hat t^4 + \hat u^4} \,\frac{{\cal C}[w_4^{hh}\,h_1^{\perp\,g}h_1^{\perp\,g}]}{{\cal C}[f_1^g \, f_1^g ]}\,,
\label{eq:R4}
\end{eqnarray}
where the transverse weights are explicitly given by
\begin{eqnarray}
w_0^{hh}  & = & \frac{1}{M_p^4}\, \left[ (\bm p_{a\sT}\cdot \bm p_{b\sT})^2 - \frac{1}{2}\, \bm p_{a\sT}^2 \,\bm p_{b\sT}^2\right ]\, ,\nonumber \\
w_2^{fh} & = & \frac{1}{M_p^2}\, \left [ 2\, 
\frac{(\bm q_\sT \cdot \bm p_{b\sT})^2}{\bm q_\sT^2} -\bm p_{b\sT}^2   \right ] \,,\quad
 w_2^{hf}\,  =  \, \frac{1}{M_p^2}\,\left [ 2\, 
\frac{(\bm q_\sT \cdot \bm p_{a\sT})^2}{\bm q_\sT^2}-\bm p_{a\sT}^2   \right ]\,,\nonumber \\
w_4^{hh} & = & \frac{1}{2 M_p^4}\, \left \{ 2 \,\left [
 2 \, \frac{(\bm q_\sT\cdot \bm p_{a\sT}) (\bm q_\sT\cdot \bm p_{b\sT})   }{\bm q_\sT^2} 
  -\bm p_{a\sT} \cdot  \bm p_{b\sT} \right ]^2 - \bm p_{a\sT}^2  \bm p_{b\sT}^2 \right \}~. 
  \label{eq:weights}
\end{eqnarray}

\begin{figure}[t]
\centering
\hspace*{-0.5cm}
\includegraphics[width=5.4cm]{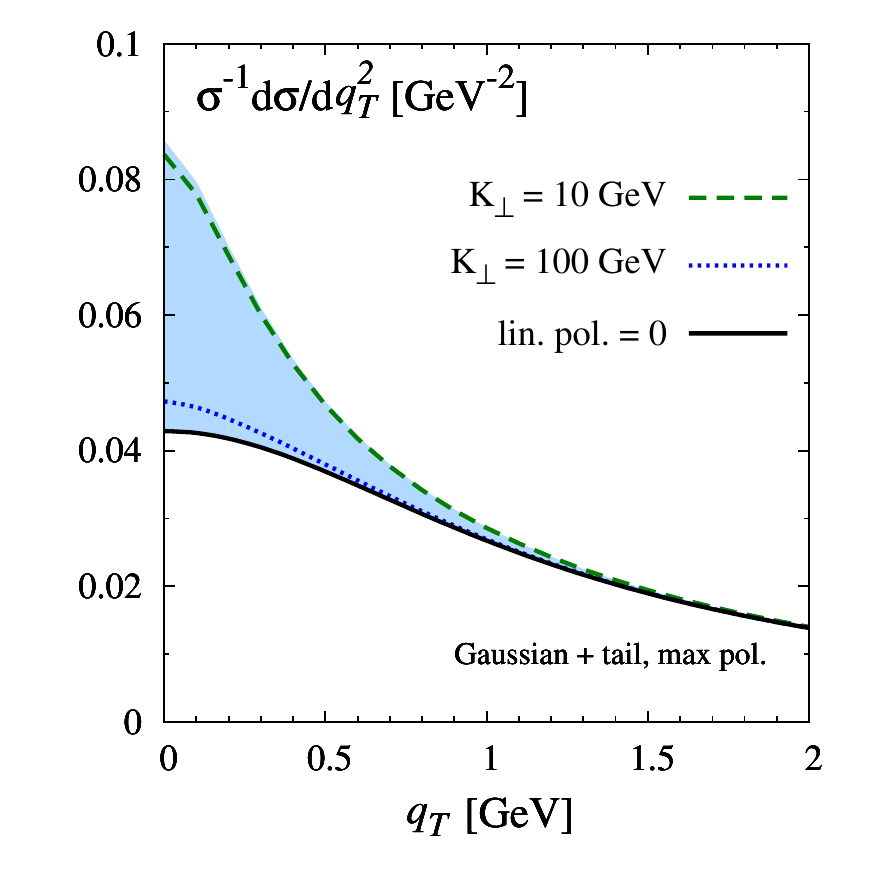}\hspace{-0.6cm}
\includegraphics[width=5.4cm]{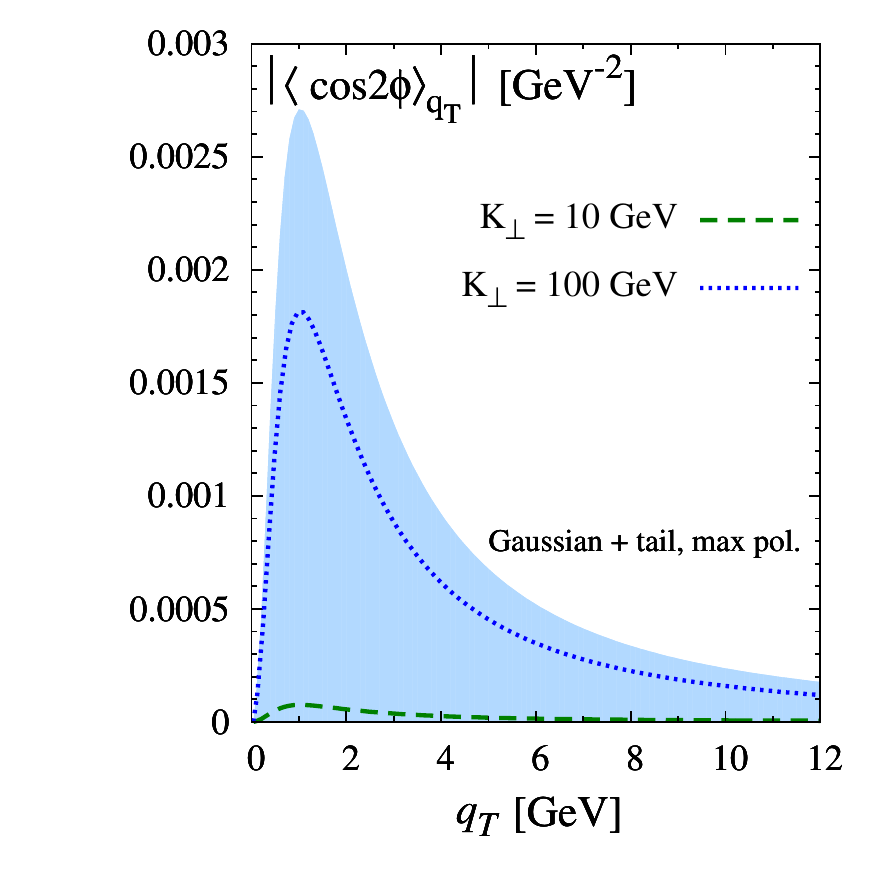}\hspace{-0.6cm}
\includegraphics[width=5.4cm]{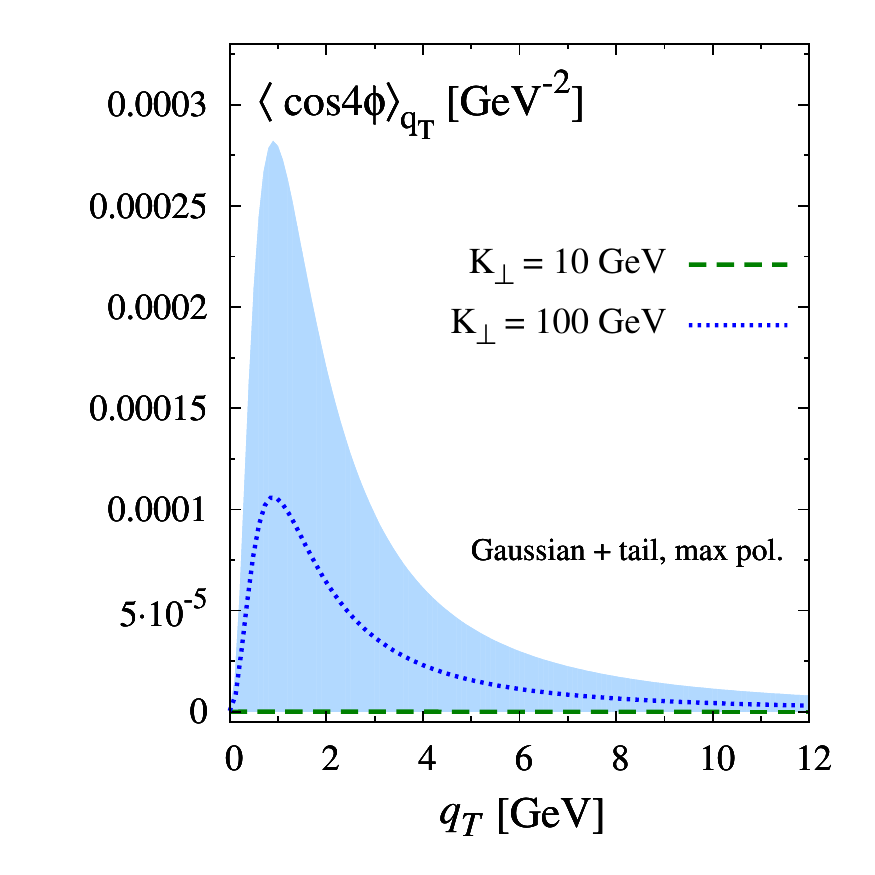}
\caption{Transverse momentum distribution (left panel), $\cos 2\phi$  and $\cos 4\phi$ azimuthal asymmetries (central and right panels) of the Higgs boson plus jet pair in the process $p\,p\to H\,{\rm jet}\, X$ for two different choices of $K_\perp$, with $\bm q^2_{\sT {\rm max}} = M^2_H/4$ and $y_H=y_{\rm j}$.  The solid line indicates the $H+$jet spectrum in absence of linear polarization. The shaded blue regions represent the range of the three observables as $K_\perp$ varies from $0$ to $\infty$.}
\label{fig:qT_db}
\end{figure}

By defining the observables~\cite{Dunnen:2014eta}
\begin{equation}
\langle \cos n\phi \rangle_{q_\sT} \equiv 
\frac{\int_0^{2 \pi} \d \phi \,\cos n\phi\, 
\d\sigma}{\sigma}\,,\qquad n=0, 2, 4\,,
\label{eq:cosnphiqT}
\end{equation}
such that their integrals over $\bm q_{\sT}^2$ give the average values of $\cos n\phi$, with $n=0, 2, 4$,
\begin{equation}
\langle \cos n\phi \rangle \equiv \frac{\int_0^{q_{\sT \rm max }^2} \d\bm q_{\sT}^2\int_0^{2 \pi} \d \phi \,\cos n\phi\, 
\d\sigma}{\sigma}  = \int_0^{q_{\sT \rm max }^2} \d\bm q_{\sT}^2\,\langle \cos n\phi \rangle_{q_\sT} \,,
\label{eq:cosnphi}
\end{equation}
it is possible to isolate the amplitudes $1+R_0$, $R_2$, $R_4$ in Eq.~(\ref{eq:csgg-2}) and access the three different convolutions of gluon TMDs. One  obtains
\begin{eqnarray}
\frac{1}{\sigma}\,\frac{\d\sigma}{\d \bm q_{\sT}^2} & \equiv & \langle 1 \rangle_{q_\sT} \,  = \,  \sigma_0(\bm q_\sT^2)\, [ 1+ R_0(\bm q_\sT^2)]\,,
\label{eq:qT} \\
\langle \cos 2\phi \rangle_{q_\sT} & = &  \frac{1}{2}\,
\sigma_0(\bm q_{\sT}^2)\, R_2(\bm q_{\sT}^2)  \,, \\
\langle \cos 4\phi \rangle_{q_\sT} & = &   \frac{1}{2}\,
\sigma_0(\bm q_{\sT}^2)\, R_4(\bm q_{\sT}^2).
\label{eq:cos4phi}
\end{eqnarray}
Predictions for these transverse momentum dependent quantities are presented in Fig.~\ref{fig:qT_db} in the specific configuration in which the Higgs boson and the jet have the same rapidities, $y_H=y_{\rm j}$, and for two different values of $K_\perp \equiv  \vert \bm K_\perp\vert $, namely 10 and 100 GeV. The  so far unknown unpolarized gluon TMD has been assumed to have the following form~\cite{Boer:2014tka},
\begin{equation}
f_1^g(x,\bm p_\sT^2) = {f_1^g(x)}\, \frac{R^2}{2\,\pi}\, \frac{1}{1 + \bm p_\sT^2\,R^2}\,,
\label{eq:f1gdb}
\end{equation}
where  $R= 2$ GeV$^{-1}$ and $f_1^g(x)$ is the unpolarized gluon distribution, integrated over its transverse momentum. In order to show the maximal effects of gluon polarization,
$h_1^{\perp \,g}$ is taken to be positive and saturating the bound in Eq.~(\ref{eq:bound}). Moreover,  $\bm q^2_{\sT \rm max } = M_H^2/4$~\cite{Boer:2014lka}. Note that $\langle \cos 2 \phi\rangle_{q_\sT}$ is the only observable sensitive to the sign of $h_1^{\perp\,g}$, and it is expected to be negative if $h_1^{\perp\,g} > 0$. It is found that $\vert\langle \cos 2 \phi\rangle\vert \approx 12\%$ at $K_\perp = 100$ GeV, while it is about 0.5\% at $K_\perp = 10$ GeV.  Furthermore, $\langle \cos 4 \phi\rangle \approx 0.2 \%$ if $K_\perp = 100$ GeV and completely negligible at $K_\perp = 10$ GeV. Analogous results are obtained if one considers a Gaussian model for $f_1^g$ and maximal gluon polarization~\cite{Boer:2014lka}.  

To conclude this section, it is important to mention that the proposed measurements require high resolution of the transverse momentum of both the Higgs boson and the jet, because one would need several bins in $q_\sT$ in the kinematic region $ q_\sT  \le 10$ GeV. In addition, the knowledge of how well the jet axis coincides with the direction of the fragmenting parton is needed.

\section{Quarkonium production in association with a photon in $pp$ collisions}

The calculation of the cross section for the process  $p(P_A)+p(P_B)\to {\cal Q}(K_{\cal Q}) + \gamma (K_{\gamma})+ X$, where ${\cal Q}$ is a $C=-$ quarkonium ($J/\psi$ or $\Upsilon$) with mass $M_{\cal Q}$, proceeds along the same lines of Higgs plus jet production presented in the previous section.  As before, the imbalance of the quarkonium-photon  pair $\qT = \bm K_{{\cal Q}\perp}+ \bm K_{\gamma\perp}$ is small, but not the individual transverse momenta $\bm K_{{\cal Q}\perp}$  and $ \bm K_{\gamma\perp}$. Therefore no forward detector is needed for the proposed measurements, in contrast to the the inclusive production of $C=+$ quarkonia that can also be used to probe gluon TMDs~\cite{Boer:2012bt,Ma:2012hh,Brodsky:2012vg}. 

\begin{figure*}[t]
\centering
{\includegraphics[width=0.31\textwidth]{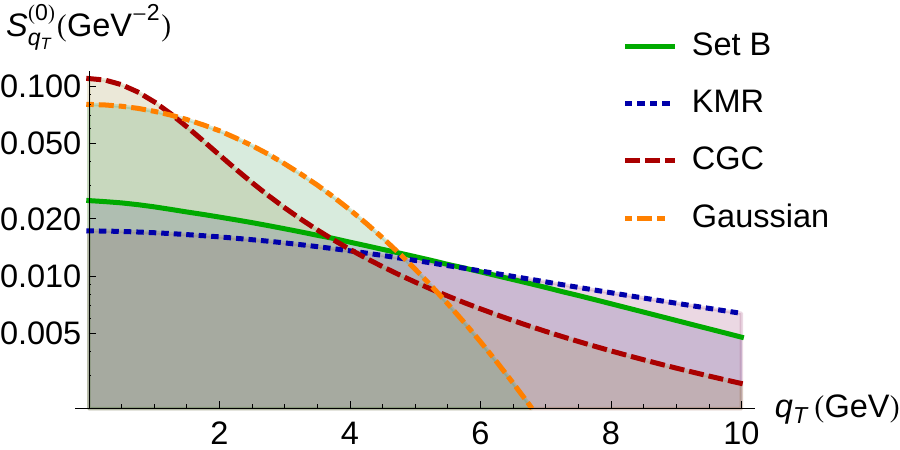}}
\hspace{0.3cm}
{\includegraphics[width=0.31\textwidth]{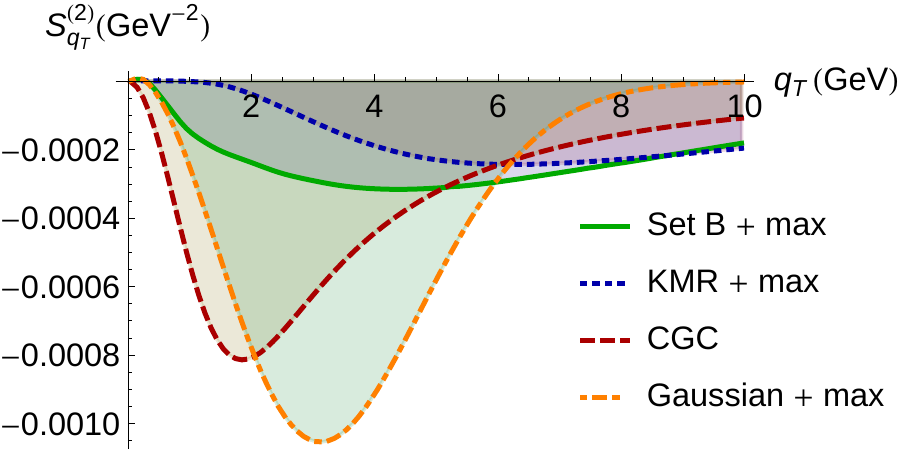}}
\hspace{0.3cm}
{\includegraphics[width=0.31\textwidth]{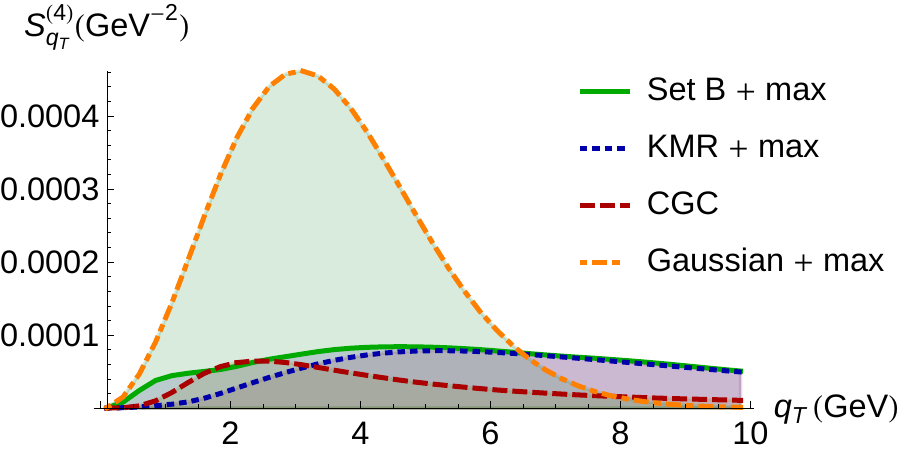}}
\caption{The TMD observables  ${\cal S}^{(0)}_{q_T}$,  ${\cal S}^{(2)}_{q_T}$ and  ${\cal S}^{(4)}_{q_T}$ for the process $p(P_A)+p(P_B)\to {\Upsilon}(P_{\cal Q}) + \gamma (P_{\gamma})+ X$  at $\sqrt{s}=14$ TeV in the kinematic region defined by  $Q=20$ GeV, $Y=0$, $\theta =\pi/2$,  and  $x_a=x_b \simeq1.4\times10^{-3}$.}
\label{fig:dsigma4dqT}
\end{figure*}

The resulting cross section $\d\sigma \equiv \d\sigma/\d Q\, \d Y\, \d^2 \qT\, \d \Omega$  has the following structure~\cite{Dunnen:2014eta},
\begin{align}\label{eq:crosssectionPsi-g}
\d\sigma &= {\cal N} \left\{
  F_1\, \mc{C} \Big[f_1^gf_1^g\Big]+ F_2 \,\mc{C} \Big[w_2^{fh} f_1^g h_1^{\perp g} + x_a\! \leftrightarrow\! x_b \Big] \cos 2\phi + F_4  \mc{C} \left[w_4^{hh} h_1^{\perp g}h_1^{\perp g}\right]\cos 4\phi \right\} \,,
\end{align}
where $Q$ and $Y$ are the invariant mass and the rapidity of the pair, to be measured, like $\bm q_{\sT}$, in the hadronic center-of-mass frame. The solid angle $\Omega=(\theta,\phi)$ is measured in the so-called Collins-Soper frame~\cite{Collins:1977iv}, defined as the frame in which the final pair is at rest and the $\hat x\hat z$-plane is spanned by $\bm P_A$ and $\bm P_B$, with the $\hat x$-axis set by their bisector. The normalization factor $\cal N$ is given by  
\begin{equation}
{\cal N} = 4 \,\alpha_s^2 \alpha \,e_Q^2\, |R_0(0)|^2\, \frac{Q^2-M_{\cal Q}^2}{3\,s \,Q^3\,M_{\cal Q}^3\,D}\,,
\end{equation}
where $R_0(0)$ is the radial wave function of the quarkonium evaluated at the origin, $e_Q$ is  the heavy quark charge and 
\begin{equation}
 D = \left[ (\alpha ^2+1)^2-(\alpha ^2-1)^2 \cos^2 \theta\right ]^2\,,
\end{equation}
with $\alpha \equiv Q/M_{\cal Q}$.
The convolutions of TMDs are defined in Eq.~(\ref{eq:Conv}), with the light-cone momentum fractions $x_{a/b} =  \exp[\pm Y]\, Q/\sqrt{s}$, the transverse weights are given in Eqs.~(\ref{eq:weights}) and the $F_{1,2,4}$ terms read
\begin{align}
 F_1 &=  1 + 2 \alpha ^2 + 9 \alpha ^4 + (6 \alpha ^4-2) \cos^2\theta + (\alpha ^2-1)^2 \cos^4\theta \,,
 \nonumber\\
 F_2 &=  -8\, \alpha^2\, \sin^2\theta,\ \qquad F_4 =  \frac{1}{2}\,(\alpha ^2-1)^2 \sin^4\theta\,~.
\end{align}
Note that in the derivation of Eq.~(\ref{eq:crosssectionPsi-g}), the Color Singlet Model~\cite{Baier:1983va} for the quarkonium production mechanism has been adopted. Color octet contributions~\cite{Bodwin:1994jh}, which could potentially lead to a breakdown of TMD factorization, are expected to be small, especially for $\Upsilon-\gamma$ production in the kinematic region described in the caption of Fig.~\ref{fig:dsigma4dqT}~\cite{Dunnen:2014eta}.

As in the previous section, the three terms in Eq.~(\ref{eq:crosssectionPsi-g}) can be singled out by means of the observables  ${\cal S}^{(n)}_{q_T} \equiv \langle \cos n\phi\rangle_{q_\sT}$, with  $n=0,2,4$, defined in Eq.~(\ref{eq:cosnphiqT}), where the cross section in the denominator is integrated over $\phi$ and $\bm q^2_\sT$, up to a value $\bm q^2_{\sT \rm{max}}\ll Q^2$. One finds
\begin{align}\label{eq:qTdistrs}
{\cal S}^{(0)}_{q_T} =\frac{\mc{C}[f_1^g f_1^g]}
 {\int  \d \bm q_\sT^2\, \mc{C}[f_1^g f_1^g]},\quad
  {\cal S}^{(2)}_{q_T}= 
  \frac{F_2\, \mc{C}[w_2^{fh} f_1^g h_1^{\perp g} + x_a \leftrightarrow x_b]}
  {2 F_1 \int  \d \bm q_\sT^2\, \mc{C}[f_1^g f_1^g]},\quad 
{\cal S}^{(4)}_{q_T}\!=  
 \frac{F_4\, \mc{C}[w_4^{hh} h_1^{\perp g} h_1^{\perp g}]}
  {2 F_1 \int \d \bm q_\sT^2\, \mc{C}[f_1^g f_1^g]}~.
\end{align}
Numerical estimates for these observables are presented in Fig.~\ref{fig:dsigma4dqT} for $\Upsilon+\gamma$ production, where $\bm q^2_{\sT \rm max}= Q^2/4$. In contrast to Higgs plus photon production, $h_1^{\perp\,g}$ does not contribute to the spectrum of the quarkonium-photon pair. Therefore, through a measurement of  ${\cal S}^{(0)}_{q_T}$, which turns out to be sizable, one could determine the shape of $f_1^g$ as a function of $q_\sT$. Furthermore,  since ${\cal S}^{(2)}_{q_T}$ and ${\cal S}^{(4)}_{q_T}$ are rather small, it would be needed to integrate them over $\bm q_\sT^2$, {\it e.g.}\ up to $Q^2/4$, to get at least an experimental evidence of a nonzero linear polarization of gluons~\cite{Dunnen:2014eta}.

\section{Summary and conclusions} 

The unpolarized and linearly polarized gluon TMDs of an unpolarized proton could be probed  in future EIC experiments by looking at the transverse momentum distributions and azimuthal asymmetries of heavy quark and jet pairs. At the LHC complementary information can be gathered by similar analyses for processes like $pp\to H\,{\rm jet}\,X$, $pp\to J/\psi(\Upsilon)\,\gamma\,X$, presented in detail in this contribution, or like $pp\to\gamma\,\gamma\,X$, proposed in Ref.~\cite{Qiu:2011ai} specifically for RHIC and for which factorization breaking terms should be absent. 

Although TMDs are not universal, their process dependence can be calculated. For instance,  despite the different gauge link structures, it has been found that in $ep \to e^\prime Q\bar Q\,X$ and in all the processes with a colorless final state, like $pp \to \gamma\, \gamma \,X$, $pp \to  H\,X$ and  $pp \to  \eta_{c,b}\,X$,
one  always probes the same \emph{effective} distribution of linearly polarized gluons, given by the sum of two of the five universal $h_1^{\perp\,g}$ functions ~\cite{Pisano:2013cya,Buffing:2013kca}. This restricted universality needs to be tested experimentally, using LHC or RHIC data. On the other hand, the gauge link 
structure for $pp\to H\,{\rm jet}\,X$ is much more complicated and still has to be investigated. It is therefore of great interest to compare the extractions of gluon TMDs from different processes, in order to learn about their process and energy scale dependences, as well as the size and importance of possible factorization breaking effects. 
 
\acknowledgments
I would like to thank Dani\"el Boer, Stan Brodsky, Maarten Buffing, Wilco den Dunnen, Jean-Philippe Lansberg, Piet Mulders, and Marc Schlegel for fruitful collaborations on this topic. 
This work was supported by the Fonds Wetenschappelijk Onderzoek - Vlaanderen (FWO) through a Pegasus Marie Curie Fellowship.

\end{document}